# Determining the Range of an Artificial Satellite Using its Observed Trigonometric Parallax

*by Michael A. Earl, Ottawa Centre Meeting Chair (earlm@sympatico.ca)*

O bserving artificial satellites is a relatively new and unique branch of astronomy that is very interesting and dynamic. One specific aspect of observing these objects is that although they appear amongst the celestial background, as deep-sky objects do, their apparent locations amongst this background depend on where you are standing on Earth at a given time. This effect is known as parallax.

When a satellite is observed at a specific time from a specific location, the satellite's equatorial coordinates can be determined using astrometric means. Its range from the observer, however, is still unknown unless the observer knows the satellite's precise orbit elements or has easy access to a radar station. However, when two or more observers, separated by some distance, observe the same satellite at the same time, their observations can be used to determine the range of the satellite using the satellite's observed trigonometric parallax.

## THE PARALLAX EXPERIMENT – THEORY

Most man-made satellites orbit the Earth at ranges from 200 to 40,000 kilometres above the Earth's surface. Because artificial satellites are so much closer to us than most of the objects we observe in the night sky, these objects will appear to be seen at different locations amongst the stellar background from different locations on the Earth. In other words, when a specific satellite at a specific time is seen at specific equatorial coordinates (Right Ascension and Declination) by one observer, another observer at another location will see it at different equatorial coordinates at that same time. The angle between the observed coordinates will depend on the distance between the observers and the range (distance) of the satellite from the observers. The parallax effect can be seen for any Earth-orbiting satellite by using two telescopes located as close together as the opposite ends of a city, like Ottawa.

Imagine two observers at points $P_1$ and $P_2$ that are separated a distance d from each other as illustrated in Figure 1. These two observers are simultaneously observing a satellite located at point S at ranges $R_1$ and $R_2$ from $P_1$ and $P_2$ respectively. A simple triangle drawn using these three points forms the basis of the range determination.

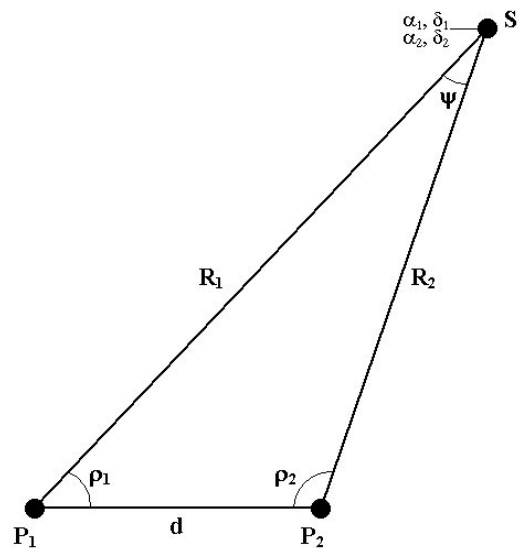

**FIGURE 1: An illustration of the parallax angle ($\psi$) that is observed for a satellite at point S at ranges $R_1$ and $R_2$ from observers $P_1$ and $P_2$, respectively, on the surface of the Earth a distance d apart.**

Now imagine that both observers record the satellite's position at exactly the same time. Observer $P_1$ will see satellite S at equatorial coordinates $\alpha_1, \delta_1$, while observer $P_2$ will see the same satellite at equatorial coordinates $\alpha_2, \delta_2$. The parallax angle ($\psi$) is determined by using the observed equatorial coordinates as shown in Eq1.

$$\cos\psi = \sin\delta_1\sin\delta_2 + \cos\delta_1\cos\delta_2\cos(\alpha_1 - \alpha_2) \quad \text{Eq1}$$

where $\psi$ = the satellite parallax angle
$\alpha_1$ = the satellite's Right Ascension observed by $P_1$
$\delta_1$ = the satellite's Declination observed by $P_1$
$\alpha_2$ = the satellite's Right Ascension observed by $P_2$
$\delta_2$ = the satellite's Declination observed by $P_2$

The range of the satellite from both observers $P_1$ and $P_2$ ($R_1$ and $R_2$ respectively) is determined by using Eq2, which is simply stating the well-known sine law for the triangle shown in Figure 1.

$$R_1/\sin\rho_2 = R_2/\sin\rho_1 = d/\sin\psi \quad \text{Eq2}$$

where $R_1$ = the range of the satellite from $P_1$
$R_2$ = the range of the satellite from $P_2$
$\rho_1$ = the angle at $P_1$ subtended by S and $P_2$
$\rho_2$ = the angle at $P_2$ subtended by S and $P_1$
d = the distance between $P_1$ and $P_2$ and
$\psi$ = the observed satellite parallax angle

In order to determine the ranges $R_1$ and $R_2$, all three angles within the triangle illustrated in Figure 1 ($\rho_1$, $\rho_2$, $\psi$) and the distance d between the observers need to be known. Since the parallax angle ($\psi$) has already been determined, and the sum of all three angles in a triangle is 180°, either angle, $\rho_1$ or $\rho_2$, need to be determined in order to know all three of these angles.

To find angle $\rho_1$, it will be necessary to determine the equatorial coordinates of observer $P_2$ as seen by observer $P_1$. The assumption here is that both observers cannot see each other. How then can observer $P_1$ know where observer $P_2$ is with respect to his/her equatorial reference frame?

Fortunately, many observatories, professional and private, have an accurate knowledge of where their observatories are on the Earth's surface in the form of their latitude, longitude (and sometimes altitude) above sea level. This information can be used to determine the values still required.

## GEODETIC COORDINATES

The Earth is not a perfect sphere. Since it spins about an axis of rotation, it is slightly flattened at its poles. Its equatorial radius is 6378.14 km and its polar radius is 6356.75 km. Because of this slight difference, your local horizon will not be exactly tangent to the line drawn from the Earth's center to the your location. Instead, your local horizon will be tangent to the line drawn from a geodetic center to your location. Figure 2 illustrates the Earth spheroid and its geometry.

The coordinates of a location on Earth that are given by a survey map or a GPS receiver are generally given in geodetic coordinates. The simple difference between the geodetic latitude and geocentric latitude, assuming the spheroid Earth illustrated in Figure 2, is shown in Eq3.

$$\tan\lambda = (b^2/a^2)\tan\lambda' \quad \text{Eq3}$$

where $\lambda$ = the geocentric latitude of the location P
a = the semi-major axis of the spheroid Earth
b = the semi-minor axis of the spheroid Earth and
$\lambda'$ = the geodetic latitude of the location P

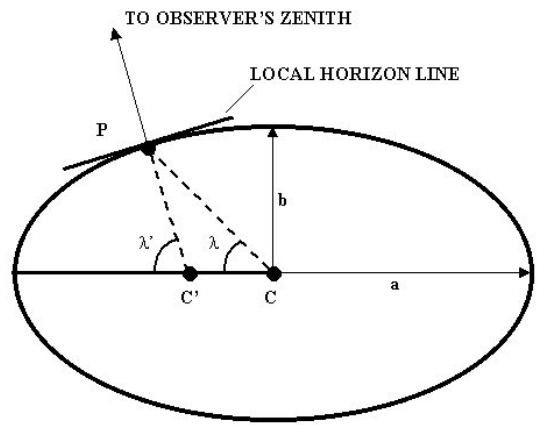

**FIGURE 2: The spheroid Earth, as illustrated by a simple ellipse. Point C depicts the geocentric center of the Earth, while point C' depicts the geodetic center as seen by the observer P. The angle $\lambda$ depicts the geocentric latitude of observer P, while $\lambda'$ depicts its geodetic latitude. The semi-major (a) and semi-minor (b) axes of the Earth are also illustrated. The eccentricity of the spheroid Earth has been exaggerated here to better accentuate the difference between the geodetic and geocentric latitude.**

Looking at Figure 3, in order to find the distance d the geocentric angle between the two locations ($\gamma$), and the distance to the two sites ($r_1$ and $r_2$) from the geocentric center of the Earth need to be determined. Eq4, the simple cosine law for triangles, shows how to determine the distance d using these values.

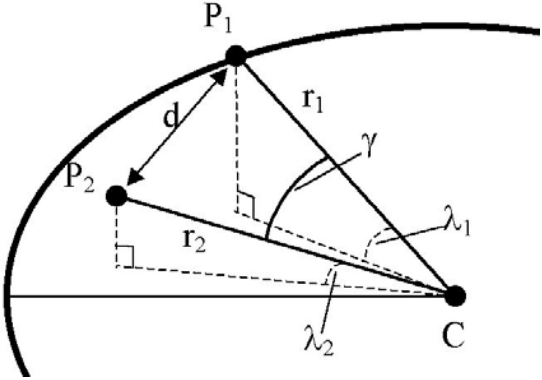

**FIGURE 3**: The determination of the distance d between two locations on Earth using their geocentric coordinates. It is assumed that observers $P_1$ and $P_2$ do not have the same longitude in this illustration.

$$d^2 = r_1^2 + r_2^2 - 2r_1r_2\cos\gamma \qquad \text{Eq4}$$

where d = distance between observers $P_1$ and $P_2$
$r_1$ = geocentric distance to observer $P_1$
$r_2$ = geocentric distance to observer $P_2$ and
$\gamma$ = geocentric angle at C subtended by $P_1$ and $P_2$

The values of $r_1$, $r_2$, and $\gamma$ now need to be determined. The angle $\gamma$ can be determined by using Eq5, which requires the geocentric longitude and latitude of both observing sites, which are already known.

$$\cos\gamma = \sin\lambda_1\sin\lambda_2 + \cos\lambda_1\cos\lambda_2\cos(\theta_1 - \theta_2) \qquad \text{Eq5}$$

where $\gamma$ = geocentric angle subtended by $P_1$ and $P_2$
$\theta_1$ = longitude of observer $P_1$
$\lambda_1$ = geocentric latitude of observer $P_1$
$\theta_2$ = longitude of observer $P_2$ and
$\lambda_2$ = geocentric latitude of observer $P_2$

Both $r_1$ and $r_2$ can be found by using the fundamental equation for ellipses in polar coordinate form, as shown in Eq6.

$$r_i^2 = [\cos^2\lambda_i/a^2 + \sin^2\lambda_i/b^2]^{-1} \qquad \text{Eq6}$$

where $r_i$ = geocentric distance to observer $P_i$
$\lambda_i$ = geocentric latitude of observer $P_i$
a = the semi-major axis of the spheroid Earth
b = the semi-minor axis of the spheroid Earth and
i = 1, 2

Determining the angle $\rho_1$ is more difficult mainly because it requires a coordinate translation from the center of the Earth to the observing site $P_1$.

**COORDINATE TRANSLATION**

Many astronomers use the equatorial coordinate system (Right Ascension and Declination) to locate objects in the night sky. The equatorial coordinate system was originally defined with its center at the geocentric center of the Earth. Most of the objects observed are so far away that the location on Earth is not a consideration when observing them. However, when the object is not far away (such as an artificial satellite) changing your observing location on Earth will cause the object to appear to shift its position amongst the stellar background, thus changing its apparent equatorial coordinates. In other words, in most cases, the parallax is negligible for planets and deep sky objects, but is significant for artificial satellites that are residing much closer to the Earth.

In order to locate an artificial satellite, simply using the geocentric Right Ascension and Declination of the satellite will not work, since no one can observe the satellite from the center of the Earth. It becomes necessary to relocate the center of the equatorial reference frame to the observer's location itself. That way, the equatorial coordinates of the satellite will be with respect to the observer's reference frame and not one that is about 6365km away at the center of the Earth. In other words, the center of the equatorial reference frame must be translated to the observer's location. Figure 4 illustrates a coordinate translation from the geocentric center of the Earth (point C) to observer $P_1$ in order to determine observer $P_2$'s apparent equatorial coordinates with respect to observer $P_1$.

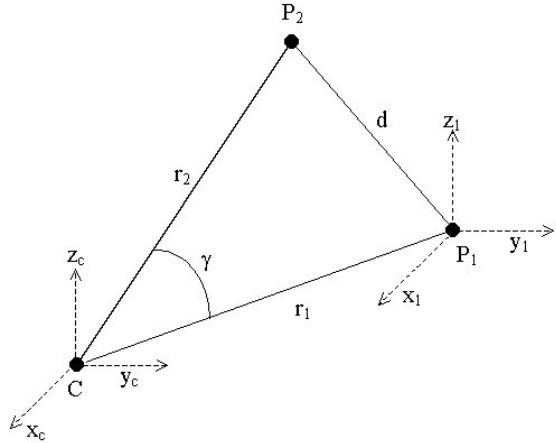

**FIGURE 4: The coordinate translation of the apparent equatorial coordinates of observer $P_2$ from the geocentric reference frame at point C to the topocentric reference frame of observer $P_1$.**

Looking at Figure 4, the coordinate translation involves the apparent Cartesian coordinates (x, y, z) of observer $P_2$. Cartesian coordinates are required because point C and point $P_1$ are related by the distance $r_1$, which is also the distance that separates the origins of their two reference frames.

The Cartesian coordinate system can also be expressed as equatorial in the following way. The x-axis is directed toward the First Point of Aries, where the Right Ascension coordinate is 0. The y-axis is directed toward the Right Ascension coordinate of 6 hours (or +90 degrees). Finally, the z-axis is directed toward the North Celestial Pole (NCP), where the Declination equals +90 degrees (this is currently ¾ of a degree from the star Polaris). This way, the Cartesian x and y-axes lie within a plane on, or parallel to, the equatorial plane of the Earth, and the z-axis coincides with, or is parallel to, the rotation axis of the Earth.

As Figure 4 indicates, the x, y, and z-axes of the Cartesian reference frames of both point C and point $P_1$ are parallel to each other. As a result, their axes will point to the same reference points, which are located at an infinite distance from Earth.

To determine the Cartesian equatorial coordinates required to perform the coordinate translation, the equatorial coordinates (R.A. and Dec.) of both observers $P_1$ and $P_2$ with respect to the geocentric reference frame at point C need to be determined. The locations of both $P_1$ and $P_2$ are already expressed as both geocentric latitude and longitude. A coordinate transformation between geocentric longitude/latitude to geocentric equatorial coordinates could be done by using the angle difference between the Prime Meridian of the Earth and the First Point of Aries, but there is an easier method. Looking at Figure 5, you will notice that a line drawn from the center of the Earth through the observer's location points approximately to the observer's zenith, but exactly along the observer's meridian. The convenient fact about the meridian is that at any time of day, it corresponds exactly to the observer's sidereal time. In other words, the sidereal time is simply that Right Ascension that lies on the observer's meridian at a specific time. So, to make a long story short, the Right Ascension of a location on the surface of the Earth as seen by the center of the Earth is precisely the surface location's sidereal time.

The Declination of an observer on the surface of the Earth with respect to the geocentric center of the Earth is exactly the observer's geocentric latitude.

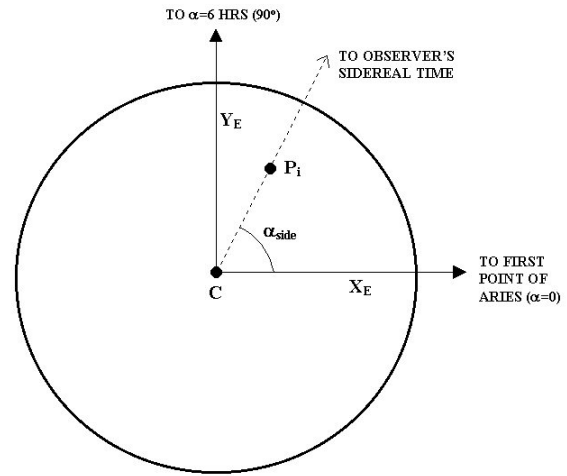

**FIGURE 5: The geocentric Right Ascension coordinate of a location on Earth is that location's local sidereal time. This illustration depicts the Earth as seen from above its northern pole.**

Now that both observers $P_1$ and $P_2$ have their positions expressed as geocentric equatorial coordinates, they can be expressed in Cartesian equatorial coordinates using Eq7 through to Eq12. The negative signs in Eq7 to Eq9 are required because the Cartesian equatorial coordinates of the center of the Earth as seen by observer $P_1$ is simply the negative of the Cartesian equatorial coordinates of observer $P_1$ as seen by the center of the Earth.

$x_{1C} = -r_1\cos\lambda_1\cos\alpha_{side1}$ **Eq7**
$y_{1C} = -r_1\cos\lambda_1\sin\alpha_{side1}$ **Eq8**
$z_{1C} = -r_1\sin\lambda_1$ **Eq9**

$x_{C2} = r_2\cos\lambda_2\cos\alpha_{side2}$ **Eq10**
$y_{C2} = r_2\cos\lambda_2\sin\alpha_{side2}$ **Eq11**
$z_{C2} = r_2\sin\lambda_2$ **Eq12**

where $x_{1C}$ = x coordinate of point C from point $P_1$
$y_{1C}$ = y coordinate of point C from point $P_1$
$z_{1C}$ = z coordinate of point C from point $P_1$
$x_{C2}$ = x coordinate of point $P_2$ from point C
$y_{C2}$ = y coordinate of point $P_2$ from point C
$z_{C2}$ = z coordinate of point $P_2$ from point C
$r_1$ = distance from point C to point $P_1$
$r_2$ = distance from point C to point $P_2$
$\lambda_1$ = geocentric latitude of observer $P_1$
$\lambda_2$ = geocentric latitude of observer $P_2$
$\alpha_{Side1}$ = sidereal time of observer $P_1$ and
$\alpha_{Side2}$ = sidereal time of observer $P_2$

The coordinate translation can now be performed using Eq13 to Eq15.

$x_{12} = x_{C2} + x_{1C}$ **Eq13**
$y_{12} = y_{C2} + y_{1C}$ **Eq14**
$z_{12} = z_{C2} + z_{1C}$ **Eq15**

where $x_{12}$ = x coordinate of point $P_2$ from point $P_1$
$y_{12}$ = y coordinate of point $P_2$ from point $P_1$ and
$z_{12}$ = z coordinate of point $P_2$ from point $P_1$

The equatorial coordinates of observer $P_2$ with respect to observer $P_1$ can now be determined using Eq16 and Eq17. Note that Eq16 has special conditions that are required to determine the correct Right Ascension quadrant. In this case, the signs of the x and y coordinate of observer $P_2$ from observer $P_1$ are required to be correct.

$\alpha_{12} = \tan^{-1}(y_{12} / x_{12})$ **Eq16**

where $\alpha_{12}$ = Right Ascension of $P_2$ as seen by $P_1$
$y_{12}$ = y coordinate of point $P_2$ from point $P_1$ and
$x_{12}$ = x coordinate of point $P_2$ from point $P_1$

**Conditions for Eq16:**

**If $x_{12} > 0$ and $y_{12} > 0$ then leave $\alpha_{12}$ as-is**
**If $x_{12} > 0$ and $y_{12} < 0$ then add 360º to $\alpha_{12}$**
**If $x_{12} < 0$ then subtract $\alpha_{12}$ from 180º**

$\delta_{12} = \sin^{-1}[z_{12} / d]$ **Eq17**

where $\delta_{12}$ = **Declination of $P_2$ as seen by $P_1$**
$z_{12}$ = **z coordinate of point $P_2$ from point $P_1$ and**
d = **the distance between observers $P_1$ and $P_2$**

The angle $\rho_1$ can now be determined using observer $P_1$'s observed equatorial coordinates of both the satellite and the other observer $P_2$ as shown in Figure 1 and Eq18. The angle $\rho_2$ can be found by using the angles $\rho_1$ and $\psi$ (which was already determined with Eq1) by using the simple triangle angle relation shown in Eq19.

$\cos\rho_1 = \sin\delta_1\sin\delta_{12} + \cos\delta_1\cos\delta_{12}\cos(\alpha_1 - \alpha_{12})$ **Eq18**

where $\rho_1$ = **the angle at $P_1$ subtended by S and $P_2$**
$\alpha_1$ = **the satellite's Right Ascension observed by $P_1$**
$\delta_1$ = **the satellite's Declination observed by $P_1$**
$\alpha_{12}$ = **Right Ascension of $P_2$ as seen by $P_1$ and**
$\delta_{12}$ = **Declination of $P_2$ as seen by $P_1$**

$\rho_2 = 180º - \psi - \rho_1$ **Eq19**

where $\rho_2$ = **the angle at $P_2$ subtended by S and $P_1$**
$\psi$ = **the satellite parallax angle and**
$\rho_1$ = **the angle at $P_1$ subtended by S and $P_2$**

Now that all three angles and the distance between the observers have been determined, the ranges of the satellite from both observers can finally be determined using Eq2.

## THE PARALLAX EXPERIMENT – PRACTICE

Of course, theory is fine, but to know the truth about the accuracies and best conditions in which to use this method, an actual satellite needs to be observed by two actual observing sites. The following two sites were chosen because they both could be controlled remotely via the Internet.

### CASTOR II

The Canadian Automated Small Telescope for Orbital Research II (CASTOR II) facility is being designed to study the orbits and characteristics of Earth-orbiting artificial satellites and to assist with the search for missing satellites. It consists of an 11-inch aperture Celestron NexStar 11 GPS telescope, and an SBIG ST-9XE CCD camera.

CASTOR II is located in Orleans at Ottawa's East end.

**SMARTSCOPE**

The SMARTScope facility is currently being developed by the RASC Ottawa Centre as a tool for public outreach and the advancement of astronomy awareness. It currently consists of a 16-inch aperture Meade LX-200 telescope, a Paramount GT-1100ME robotic mount, an Apogee AP7p CCD camera, and a HomeDome observatory dome. At present, the facility is thoroughly being tested as a fully remotely controlled observatory for the general public.

SMARTScope is located on the grounds of the Communications Research Centre (CRC) in Shirley's Bay at Ottawa's West end.

Although SMARTScope was not specifically designed for observing artificial satellites, the satellite parallax experiment was conducted with the help of this facility as one such method of testing its remote control reliability and capabilities.

**HOW THE EXPERIMENT WAS CONDUCTED**

The choice of the test satellite was conducted by considering several criteria. The satellite could not be seen near the horizon of either observation site, as a high atmospheric refraction would certainly taint the results. Since the observing sites chosen are mostly East-West in separation, the best satellite would be located near the northern or southern meridian in order to maximize the observed parallax angle. In order to minimize the timing errors, the satellite would need to be at a large enough distance away such that its apparent angular velocity is small. In order to verify that the timing was as close to simultaneous as possible, the satellite had to be a quick tumbler, therefore inactive, and exhibit bright enough reflections such that the tumble rate could be easily seen.

The satellite chosen was the inactive Russian communications satellite Molniya 3-39 (#20813 in the NORAD catalogue of satellites). Previous observations of the satellite done by the author have determined that its tumble period of 3.45 seconds per revolution has not changed appreciably in several years. The Molniya satellites also have a very variable range from the observing location due to its high eccentricity of orbit. Having an orbital inclination of about 63.5 degrees (nearly a polar orbit), it can easily appear in the northern or southern sky, which can satisfy the "near the meridian" criteria stated earlier.

The CASTOR II facility was operated locally from its location in Orleans. The SMARTScope facility was operated remotely from CASTOR II's location using the Virtual Network Computing (VNC) software. Each facility was controlled by its own computer so that timing errors would be kept to a minimum. It was hoped at the time that the cameras would be instructed to open at nearly the same time so that both images would be taken nearly simultaneously. Since the satellite would be quickly tumbling during the exposure time, the simultaneity of the exposure times could be checked using the "flashes" observed within the satellite streaks obtained.

After previously using both CCD cameras corresponding to the two sites, it was determined that the shutter opening time of both cameras corresponded to approximately 0.5 seconds after the command to open the CCD shutters had been sent. Therefore an additional 0.5 seconds had to be added to each of the time tags indicated on the resultant Flexible Image Transport System (FITS) images.

**THE IMAGES**

The two images of the artificial satellite Molniya 3-39, corresponding to each of the two sites, are shown in Figure 6 (for CASTOR II) and Figure 7 (for SMARTScope). A combined image, using the images from each site, is shown in Figure 8.

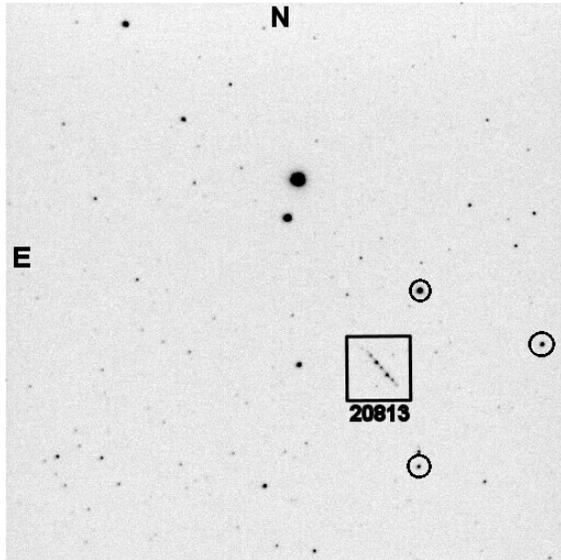 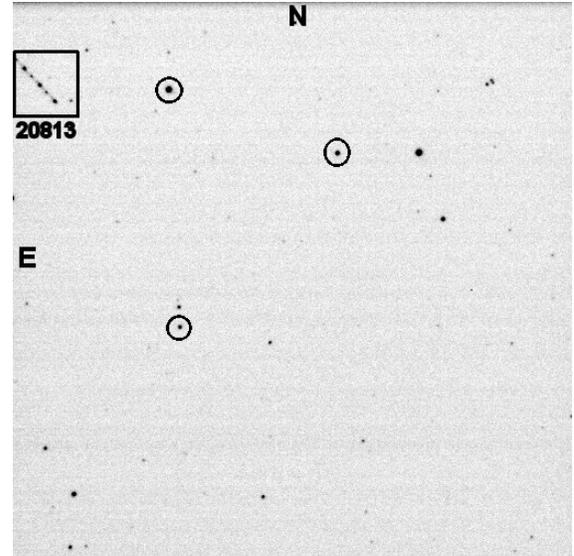

**FIGURE 6:** Artificial satellite Molniya 3-39 (#20813) as seen by the CASTOR II facility at 05:10:35.5 U.T.C. December 8, 2003. The satellite streak is within the black box. The quick tumbling of the satellite in space created the dotted pattern of the streak. The travel of the satellite was right to left, increasing in Right Ascension. Three reference stars are also indicated within the black circles. The exposure time was 10 seconds. The field of view is 13.25 by 13.25 minutes of arc. The negative of the actual image was used to better show the satellite streak.

**FIGURE 7:** Artificial satellite Molniya 3-39 (#20813) as seen by the SMARTScope facility at 05:10:35.5 U.T.C. December 8, 2003. The satellite streak is within the black box. The quick tumbling of the satellite in space created the dotted pattern of the streak. The same three reference stars identified in Figure 6 are within the black circles in this figure. The exposure time was 10 seconds. The field of view is 10.0 by 10.0 minutes of arc. The streak looks slightly longer in this image compared to CASTOR II's image, despite the identical exposure times, because of the smaller field of view of the SMARTScope detector. The negative of the actual image was used to better show the satellite streak.

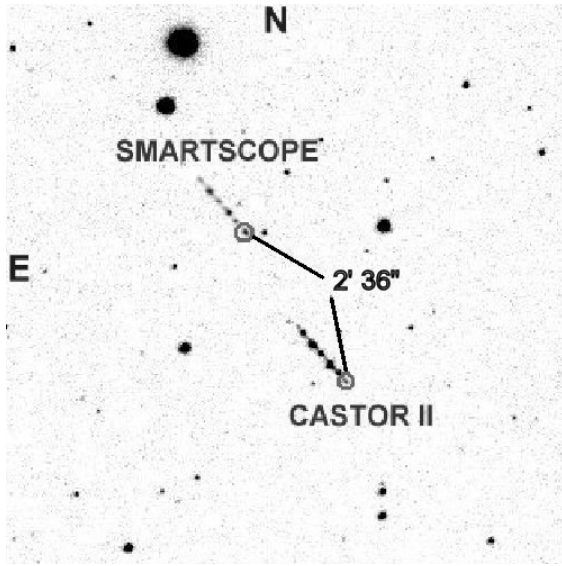

**FIGURE 8:** The apparent positions of the Molniya 3-39 satellite as seen by both the CASTOR II and SMARTScope facilities. The SMARTScope image of the satellite was grafted onto the CASTOR II image and adjusted to the CASTOR II image scale. The original CASTOR II image was cropped to better show the satellite streaks. The endpoints of both relative satellite streaks are indicated within the black circles. These endpoints were used to determine the parallax angle of the satellite, which is also indicated on this image.

Looking at Figure 8, the highlighted endpoints of the CASTOR II and SMARTScope streaks seem to indicate that the satellite was exhibiting a flash (bright reflection) at the time both shutters had opened. The CASTOR II image indicates that the flash was near its end when CASTOR II's CCD shutter opened, while the SMARTScope image indicates that the flash had nearly began when SMARTScope's CCD shutter opened. Looking at the general satellite streak itself, the duration of the flashes were very small compared to the overall exposure time of 10 seconds, so the difference in time between the CASTOR II and SMARTScope images was certainly small enough to render this portion of the timing error negligible. The possibility that both images depict adjacent flashes (3.45 seconds apart) is not possible, since the time tags of both images indicated that the "shutter open" commands were received nearly simultaneously.

## THE RESULTS

The collected data from the CASTOR II and SMARTScope facilities was analyzed using the equations stated in the Theory section of this paper. Several other images were taken which yielded similar results to those shown below.

Some numbers seem to have greater significant figures than they deserve. The large number of decimal places was preserved until the end to avoid accumulating rounding errors.

**OBSERVER 1 ($P_1$) = CASTOR II**
**OBSERVER 2 ($P_2$) = SMARTSCOPE**

**TEST SATELLITE = MOLNIYA 3-39 (#20813)**

**OBSERVATION TIME = 05:10:35.5 U.T.C.**
**OBSERVATION DATE = DECEMBER 8, 2003**

The geodetic coordinates of both sites were recorded. A right-hand-rule convention was used for the longitudes to match that of the equatorial coordinate system.

$\theta_1 = -75° \ 32' \ 11" = -75°.536389$
$\lambda'_1 = +45° \ 28' \ 27" = +45°.474167$
$\theta_2 = -75° \ 53' \ 25" = -75°.890278$
$\lambda'_2 = +45° \ 21' \ 14" = +45°.353889$

Using Eq3, the geocentric latitudes of both sites were determined.

$\lambda_1 = +45° \ 16' \ 54" = +45°.281711$
$\lambda_2 = +45° \ 09' \ 41" = +45°.161425$

Using Eq5, the geocentric angle subtended by both facilities was determined.

$\gamma = 0°.276772 = 16' \ 36".38$

Astrometric analysis was performed on both the CASTOR II and SMARTScope images to determine the coordinates of the first endpoint for each case. For the best astrometric accuracy, the star catalogue used was the United States Naval Observatory A 2.0 (USNO A2.0). To obtain the angular equivalent of the Right Ascensions, the original coordinate format (h-m-s) was multiplied by 15 degrees per R.A. hour.

$\alpha_1$ (J2000.0) = $02^h \ 59^m \ 46^s.59$ = $44°.944125$
$\delta_1$ (J2000.0) = $+55° \ 06' \ 27".94$ = $+55°.107761$
$\alpha_2$ (J2000.0) = $02^h \ 59^m \ 57^s.32$ = $44°.988833$
$\delta_2$ (J2000.0) = $+55° \ 08' \ 34".45$ = $+55°.142903$

The parallax angle between the two observed equatorial coordinates of the streak endpoints was determined using Eq1.

$\psi = 0°.043456 = 2'\ 36".44$

The geocentric distances of the two facilities were then determined using Eq6.

$r_1 = 6367.312889$ km
$r_2 = 6367.357792$ km

The distance between the two facilities was determined using Eq4.

$d = 30.757932$ km

The apparent sidereal times for both facilities were determined using the observers' longitudes and the observation time entered into the United States Naval Observatory Multiyear Interactive Computer Almanac (USNO MICA) (see References section for the URL) freely available for use on the Internet.

$\alpha_{side1} = 05^h\ 14^m\ 39^s.2892 = 78°.663708$
$\alpha_{side2} = 05^h\ 13^m\ 14^s.3559 = 78°.309833$

The Cartesian equatorial coordinates of the SMARTScope facility as seen by the CASTOR II facility was determined using Eq7 to Eq15.

$x_{1C} = -880.656317$ km
$y_{1C} = 4392.771856$ km
$z_{1C} = -4524.452818$ km

$x_{C2} = 909.699373$ km
$y_{C2} = -4396.571864$ km
$z_{C2} = 4515.069008$ km

$x_{12} = 29.043056$ km
$y_{12} = -3.800000$ km
$z_{12} = -9.383810$ km

The equatorial coordinates of the SMARTScope facility as seen by the CASTOR II facility were determined using Eq16 and Eq17.

$\alpha_{12} = 352°.545752 = 23^h\ 30^m\ 10^s.98$
$\delta_{12} = -17°.763329$

In order to verify that the above equatorial coordinates were the true ones, a coordinate transformation from equatorial to Alt-Az coordinates was performed using the handy equations on page 31 of the 2004 Observer's Handbook. The Alt-Az coordinates of SMARTScope as seen by CASTOR II will not change with time as its Right Ascension will, and so they can be used as reference coordinates for any future experiments using both facilities.

$AZ_{12} = 254°.701508 = 254°\ 42'\ 05".43$
$ALT_{12} = -9°.921248 = -9°\ 55'\ 16".49$

The SMARTScope facility's geodetic coordinates are mainly West, but a little South of CASTOR II. Therefore, the SMARTScope facility should not be located exactly due West (270°) in azimuth, but slightly South as well. The Alt-Az coordinates are reasonable, so the recently determined equatorial coordinates can be used.

The angles $\rho_1$ and $\rho_2$ in Figure 1 were then determined using Eq18 and Eq19.

$\rho_1 = 85°.287476$
$\rho_2 = 94°.669068$

Finally, the range of the Molniya 3-39 satellite from both the CASTOR II and SMARTScope facilities at the time specified were determined using Eq2.

$R_1 = \underline{40\ 419\ km}$
$R_2 = \underline{40\ 417\ km}$

The "true" ranges from both facilities were determined by propagating the most up-to-date Keplerian orbit elements of the satellite provided at the time.

$R_1$ (true) = 39 023 km
$R_2$ (true) = 39 018 km

## SOURCES OF ERROR

There are two significant sources of error that could be minimized in the future.

The first significant source of error is the choice of the test satellite's apparent angular velocity. It is true that a satellite will appear to travel more slowly in the observer's sky the further away the satellite is, and therefore be less affected by timing errors. However, the further the satellite is from the observer, the smaller the parallax angle, and the more significant the error becomes for a specific baseline distance. The test satellite chosen was about 40,000km from both observing locations, however, the baseline between the sites was a mere 31km. This combination of a large satellite range and a small baseline increased the error sensitivity because of the small parallax angle observed. The results of this experiment showed just how sensitive this error could be.

The second significant source of error is certainly the streak endpoint detection error. When determining where the endpoint of a satellite streak is most likely located, factors such as the resolution of the detector, the brightness of the endpoint compared to the image background, and the brightness variability of the satellite as it travels through space will introduce errors that can be large enough to be noticeable if the overall parallax angle is small enough. CASTOR II's resolution is currently 1.56 arc-seconds per pixel, while SMARTScope's resolution is currently 1.15 arc-seconds per pixel. It is likely that CASTOR II's lower resolution did introduce some error in the determined parallax angle. However, CASTOR II's resolution as a satellite tracking facility was chosen for two reasons. One was endpoint determination accuracy, and the other was sensitivity. If the resolution of CASTOR II is too high (pixels are too small), it may not be able to see the fainter satellites because the exposure time per pixel as the satellite travels across the CCD detector would be too short. The trade-offs resulted in the necessity to sacrifice some accuracy for sensitivity, and vice-versa. Unfortunately, this sacrifice most likely resulted in the range error in this experiment. This is not to say that SMARTScope resolution did not cause any errors, but since its resolution is better, its endpoint detection errors were probably smaller.

The largest error in satellite tracking is indeed the endpoint detection error. Determining where the true apparent location of the satellite was, corresponding to the time the shutter opened and closed, can be a very difficult process. This is especially true of tumbling satellites. A tumbling satellite can appear invisible at periodic times due to its brightness being too low to overcome the sky background, interfering background stars, or even the satellite's own brightness flare-ups. If this occurs at the time the shutter is opened or closed, the endpoint will not be seen and another part of the streak could be mistaken for the true endpoint location. It is also possible that SMARTScope's higher resolution made for a dimmer streak, thereby increasing the probability that an endpoint was tagged incorrectly. The endpoint of SMARTScope's streak indicated in Figure 8 might actually have been located several arc-seconds away due to the brightness of the flash covering up the real (dim) endpoint. Since CASTOR II's streak endpoint coincided with the ending of a bright flare-up, it is less likely that CASTOR II's endpoint is inaccurate because of inaccurate endpoint detection.

So, to minimize these errors, it will be necessary to keep the test satellite range high, but increase the baseline distance between the two observing sites in order to increase the observed parallax angle. Increasing the detector's resolution may help, but if the test satellite is tumbling, or is difficult to detect in the first place, the increased resolution may ultimately increase the endpoint detection errors, and thus may defeat the intended purpose of improving ranging accuracy. This also presents a trade-off situation that could take much work to resolve. Increasing the timing accuracy by further investigating the timing offsets of the CCD shutter will reduce the need for a larger a-priori satellite range.

When all is said and done, the sources of the errors experienced in this experiment are the same as those for any satellite tracking facility. This experiment, however, may be used to fine-tune the tracking data accuracies of all the tracking facilities involved, since the parallax angle is so sensitive to tracking data errors, especially for those facilities with smaller distance baselines.

## CONCLUSIONS

I did not expect to get results that were super-accurate as to rival radar-ranging facilities. This was an initial investigation into how inaccurate this method could be given loose constraints. This is especially true of the resolution of both CASTOR II and SMARTScope facilities. They were certainly not designed to do work such as this, especially given the small baseline. This experiment verified that indeed the resolution of both systems was too low to compensate for the small observing baseline used. Better results will be experienced if the baseline is increased. If a large satellite range is maintained, the probability that both observing sites can view the same satellite at the same time is also better.

Overall, a very fun experiment to do!

## ACKNOWLEDGEMENTS

Special thanks go to: SMARTScope manager Chris Teron, and the SMARTScope team for allowing me to use the SMARTScope facility for this experiment; Lt. Col. (retd.) Phillip W. Somers, who introduced me to, and taught me most of what I know about the subject of satellite tracking.

*Michael A. Earl was the Senior Technician of the Space Surveillance, Research, and Analysis Laboratory (SSRAL) at the Royal Military College (RMC) of Canada in Kingston, Ontario from 1997 to 2001. During that time, he designed, constructed, tested, and operated the Canadian Automated Small Telescope for Orbital Research (CASTOR) satellite tracking facility. He is currently independently designing his own satellite tracking facility (CASTOR II) to be used for further research in the subject of satellite tracking. During his spare time, he is the Meeting Chair of the Ottawa chapter of the RASC and a proud member of the SMARTScope team.*